# Influence of random roughness on cantilever curvature sensitivity

O. Ergincan, G. Palasantzas,[a)] B.J. Kooi

Zernike Institute for Advanced Materials, Nijenborgh 4, 9747 AG Groningen University of Groningen, The Netherlands

**Abstract**

In this work we explore the influence of random surface roughness on the cantilever sensitivity to respond to curvature changes induced by changes in surface stress. The roughness is characterized by the out-of-plane roughness amplitude $w$, the lateral correlation length $\xi$, and the roughness or Hurst exponent $H$ ($0<H<1$). The cantilever sensitivity is found to decrease with increasing roughness (decreasing $H$ and/or increasing ratio $w/\xi$) or equivalently increasing local surface slope. Finally, analytic expressions of the cantilever sensitivity as a function of the parameters $w$, $\xi$, and $H$ are derived in order to allow direct implementation in sensing systems.

---

[a)]Corresponding author: G.Palasantzas@rug.nl

Micro/Nanoelectromechanical systems (MEMS/NEMS) are important devices that combine in many cases the advantages of mechanical systems with the speed and large scale integration of silicon based microelectronics. As a result a large number of groups are experimenting on various aspects to understand fully the properties and potential of MEMS/NEMS.[1-11] As a matter of fact, micromechanical cantilevers allow mass resolution down to femtograms in air environment.[12] The sensitivity is determined by the effective vibratory mass of the resonator (determined by geometry, configuration, and material properties of the resonant structure), and the stability of the device resonance frequency.[6]

Complete understanding of cantilever bending is crucial for ultra-sensitive applications including bending experiments to monitor the evolution of material properties, monitoring the surface stress during bending, adsorption studies, self-assembly, thin film deposition, and molecular-recognition based on-chip biomedical sensing devices.[13-22] Such experiments exploit Stoney's equation, which assumes planar geometry, to relate the cantilever bending to the magnitude of the surface stress ($f$) change $\Delta f$.[23] However, the surfaces of real cantilevers (and in more general material systems; even for the most thoroughly polished surfaces) have random surface roughness on different lateral length scales.

Recently, it was shown that the response of the curvature of cantilevers to changes in their surface stress depends significantly on the surface morphology.[23] This dependence was attributed to the transverse coupling between the out-of-plane and in-plane components of the surface-induced stress. Moreover, roughness corrections were introduced, which are highly important for experiments measuring the surface stress on nominally planar surfaces.[23] However, calculations of the roughness effects on the cantilever sensitivity as a function of characteristic parameters of random rough surfaces are still missing. This will be the topic of the present paper.

Initially we will present the theory of cantilever bending including the general roughness corrections from Ref. 23, and afterwards we will implement specifics of random self-affine roughness. It has been shown that the in-plane stress *T* in a surface layer, which is required to undo the additional in-plane strain components induced by the stress at equilibrium in the layer (if it was detached from the substrate and allowed to strain freely under the influence of the surface stress) while allowing for free relaxation along the normal is given by [23]

$$T = \frac{1}{h^L} \frac{A_{rough}}{A} \left( s_{II} - \frac{v^L}{1-v^L} s_\perp \right), \quad (1)$$

with $h^L$ the mean thickness of a layer deposited on one side of the cantilever surface (see inset Fig. 1) and having a rough surface with area $A_{rough}$, and a Poisson ratio $v^L$. We assume an isotropic and continuously differentiable rough profile *h(r)* with *r=(x, r)* the in-plane position vector ($h^L = <h(r)>$ with $<....>$ denoting an ensemble average). The stress components $s_{II}$ and $s_\perp$ are respectively the in-plane and out-of-plane components. They are given respectively by [23] $s_{II} = (A/A_{rough})\langle f(3+\cos 2\theta)/4\cos\theta \rangle$ and $s_\perp = (A/A_{rough})\langle f(\sin^2\theta/\cos\theta) \rangle$. *f* is the position dependent scalar surface stress. The angle $\theta$ (inset, Fig.1) is defined as $\tan\theta = |\nabla h|$ and A is the average flat surface area. Substitution in Eq. (1) yields for the stress *T*:

$$T = \frac{1}{2h^L} \left\langle f \left[ \frac{1+v^L}{1-v^L} \cos\theta + \frac{1-3v^L}{1-v^L} \sec\theta \right] \right\rangle \quad (2)$$

For weak roughness ($\tan\theta = |\nabla h| \ll 1$ and $h^L \gg w$ with $w = \sqrt{\langle (h-h^L)^2 \rangle}$ the rms roughness amplitude), $s_{II}$ is only weakly affected by surface roughness, while the dominant effect arises from the out-of-plane stress component $s_\perp$. Local values of $f$ and $\theta$ will be correlated due to dependence of the surface properties on the surface crystallographic orientation.[24]

Furthermore, if we assume a Gaussian height distribution of the height profile (which is reasonable in many cases of deposited overlayers and depending on the growth mode),[25-29] the rough surface area in Eq. (1) is given by [25] $A_{rough}/A = \int_0^{+\infty} du \left(\sqrt{1+\rho_{rms}^2 u}\right) e^{-u}$ with $\rho_{rms} = \langle (\nabla h)^2 \rangle^{1/2}$ the rms local surface slope. The latter in terms of Fourier-transform analysis is given by $\rho_{rms} = \left(\int_o^{Q_c} q^2 \langle |h(q)|^2 \rangle d^2q\right)^{1/2}$ and $\langle |h(q)|^2 \rangle$ the power roughness spectrum.[26] $Q_c = \pi/a_o$ is the integration limit with $a_o$ the minimum lower roughness cut-off of the order of atomic dimensions. For weak roughness ($\rho_{rms} \ll 1$) we obtain $A_{rough}/A \approx 1 + (\rho_{rms}^2/2) = 1 + \langle \theta^2 \rangle/2$ since $\rho_{rms} \approx \sqrt{\langle \theta^2 \rangle}$ ($\rho_{rms} = \sqrt{\langle |\nabla h|^2 \rangle} = \sqrt{\tan^2\theta}$). Note that the first order perturbative expansion for the surface area ratio $A_{rough}/A$ is also independent of the assumption of Gaussian height distribution.

A wide variety of surfaces exhibit a so-called self affine roughness. This type of roughness is characterized, besides the rms roughness amplitude $w$, by the lateral correlation length $\xi$ (indicating the average lateral feature size), and the roughness exponent $0<H<1$. Small values of $H\sim 0$ corresponds to jagged surfaces, while large values of $H\sim 1$ to a smooth hill valley morphology.[26-29] For self affine roughness, the power spectrum obeys the scaling behavior $<|h(q)|^2> \propto q^{-2-2H}$ if $q\xi >> 1$ and $<|h(q)|^2> \propto const.$ if $q\xi << 1$.[27,28] This scaling is satisfied by the analytic model:[28]

$$\langle |h(q)|^2 \rangle = \frac{aw^2\xi^2}{(1+q^2\xi^2)^{1+H}}. \tag{3}$$

The parameter '$a$' in Eq. (3) is obtained by the normalization condition $\int_{0<q<Q_c} q^2 \langle |h(q)|^2 \rangle d^2q = w^2$ yielding $a = (H/\pi)/[1-(1+Q_c^2\xi^2)^{-H}]$.[28] This is equivalent to the fact that the height-height correlation $C(\vec{r}) = <h(\vec{r})h(\vec{0})> = \int <|h(q)|^2> e^{-i\vec{q}\cdot\vec{r}} d^2q$ obeys the condition: $C(\vec{r}=\vec{0}) = w^2$. Furthermore, we obtain for the local slope $\rho_{rms}$ the analytic expression:

$$\rho_{rms} = \frac{w}{\xi}\left\{\frac{a\pi}{(1-H)}\left[\left(1+Q_c^2\xi^2\right)^{1-H}-1\right]-1\right\}^{1/2}. \tag{4}$$

Analytic expressions in the limiting cases $H=0$ and $1$ can be obtained using the identity $\ln(x) = \lim_{c\to 0}(1/B)(x^B - 1)$. Therefore, we obtain $\rho_{rms}|_{H=1} = (w/\xi)\{a\ln(1+Q_c^2\xi^2)-1\}^{1/2}$ with $a_{H=1} = (1/\pi)(1+(Q_c\xi)^{-2})$, and $\rho_{rms}|_{H=0} = (w/\xi)\{a\pi(Q_c\xi)^2-1\}^{1/2}$ with $a_{H=0} = 1/[\pi\ln(1+Q_c^2\xi^2)]$.

For weak roughness ($\rho_{rms} \approx \sqrt{<\theta^2>} <<1$), by considering the Taylor expansions $\cos x \cong 1-x^2/2+...$ and $\sec x \cong 1+x^2/2+...$, Eq. (2) yields after series expansion the simpler form[13] $T \cong (\langle f \rangle_S / h^L)(1 - v^L/(1-v^L)\rho_{rms}^2)$ where substitution of Eq. (4) gives the analytic form for the cantilever sensitivity, which is defined by the ratio $T/T_o$,

$$\frac{T}{T_o} \cong \left[1 - \frac{v^L}{(1-v^L)}\left(\frac{w}{\xi}\right)^2 \left\{\frac{a\pi}{(1-H)}\left[(1+Q_c^2\xi^2)^{1-H} - 1\right] - 1\right\}\right] \qquad (5)$$

with $T_o = \langle f \rangle_S / h^L$ representing the effective stress for a planar surface with the same surface stress as the rough surface.

Figure 1 shows the cantilever sensitivity $T/T_o$ as a function of the local slope $\rho_{rms}$. In fact, Eq. (5) defines a limiting value of the local slope $\rho_{rms}$ for which $T=0$, yielding $\rho_{rms}|_{max} = \sqrt{(1-v^L)/v^L}$. For Poisson ratios $v^L=0.18$ (Si(111))[30] and $v^L=0.28$ (Si(100))[30] we obtain respectively $\rho_{rms/max}=2.13$ and $\rho_{rms/max}=1.6$. For a metallic overlayer as gold (widely used to coat cantilevers) with $v^L=0.44$[30] we obtain $\rho_{rms/max}=1.12$. These are relatively significant values for $\rho_{rms}$ and the pertubative expansion of Eq. (5) are valid only for local slopes $\rho_{rms} < 1$. It is clear (e.g., from Fig. 3 and Eq. (5)) that the local surface slope influence is minimized with decreasing Poisson ratio. The cantilever sensitivity can decrease significantly with changing crystallographic structures corresponding to different Poisson ratios and the effect becomes more pronounced when the local slope increases or equivalently the surface roughness increases.

In order to obtain a direct feeling of the influence of the characteristic roughness parameters ($w, \xi, H$), we present in Fig. 2 the cantilever sensitivity $T/T_o$ as a function of the long wavelength roughness ratio $w/\xi$ for various roughness exponents $H$. The inset shows simultaneously the corresponding local slopes to ensure that our calculations are performed for $\rho_{rms} < 1$ and thus to ensure validity of the present formalism. It is clear that for lower roughness exponents $H$ and/or larger roughness ratios $w/\xi$ the cantilever sensitivity decreases rather drastically. A similar behavior is depicted in Fig. 3, where the cantilever sensitivity is depicted as a function of $w/\xi$ for two different Poisson ratios $v^L$. The inset, showing a typical gold rough surface deposited onto Si with $H=0.9$, $w= 7\ nm$, and $\xi=30\ nm$ yielding $w/\xi=0.23$, indicates that the random roughness parameters used in this study are often met in experimental coatings depending on material and preparation conditions.

In conclusion, random surface roughness influences the cantilever sensitivity $T/T_o$ to respond to changes of the associated surface stress when the cantilever sensor is used with a sensing layer on top. In more detail the cantilever sensitivity $T/T_o$ is found to decrease with increasing local surface slope or equivalently increasing surface roughness (decreasing $H$ and/or increasing ratio $w/\xi$). Even weak local surface slopes are shown to have a significant effect on the cantilever sensitivity $T/T_o$. Finally, the analytic expressions derived for the cantilever sensitivity $T/T_o$ as a function of characteristic roughness parameters allow direct implementation in sensing systems if measurements of the characteristic roughness parameters of cantilever surfaces are performed and the corresponding roughness correction is taken into account for the surface stress $T$. This is desirable since cantilever bending studies are a major source for current experimental data of surface stress $T$.

**Acknowledgments:** We would like to acknowledge financial support by the STW grant 10082.

**Figure Captions**

**Figure 1** Cantilever sensitivity $T/T_0$ as a function of local surface slope $\rho_{rms}$ with a varying correlation lengths $10\ nm \leq \xi \leq 500\ nm$, $w=1\ nm$, $H=0.5$, and two different Poisson ratios $v^L=0.18$ (corresponding to $Si(111)$) and $v^L=0.28$ (corresponding to $Si(100)$). The inset shows a random rough surface of an overlayer on a cantilever surface, $\theta$ is the inclination angle between the normal vectors n and $\hat{n}$.

**Figure 2** Cantilever sensitivity $T/T_0$ as a function of long wavelength roughness ratio $w/\xi$ for $w=1\ nm$ and $H=0.5$, and two different Poisson ratios $v^L=0.18$ (corresponding to $Si(111)$) and $v^L=0.28$ (corresponding to $Si(100)$).

**Figure 3** Cantilever sensitivity $T/T_0$ as a function of long wavelength roughness ratio $w/\xi$ for $w=1\ nm$, different roughness exponents $H$, and Poisson ratio $v^L=0.18$ (corresponding to $Si(111)$). The inset shows a typical gold rough surface deposited onto Si with $H=0.9$, $w=7\ nm$, and $\xi=30\ nm$ yielding $w/\xi=0.23$.

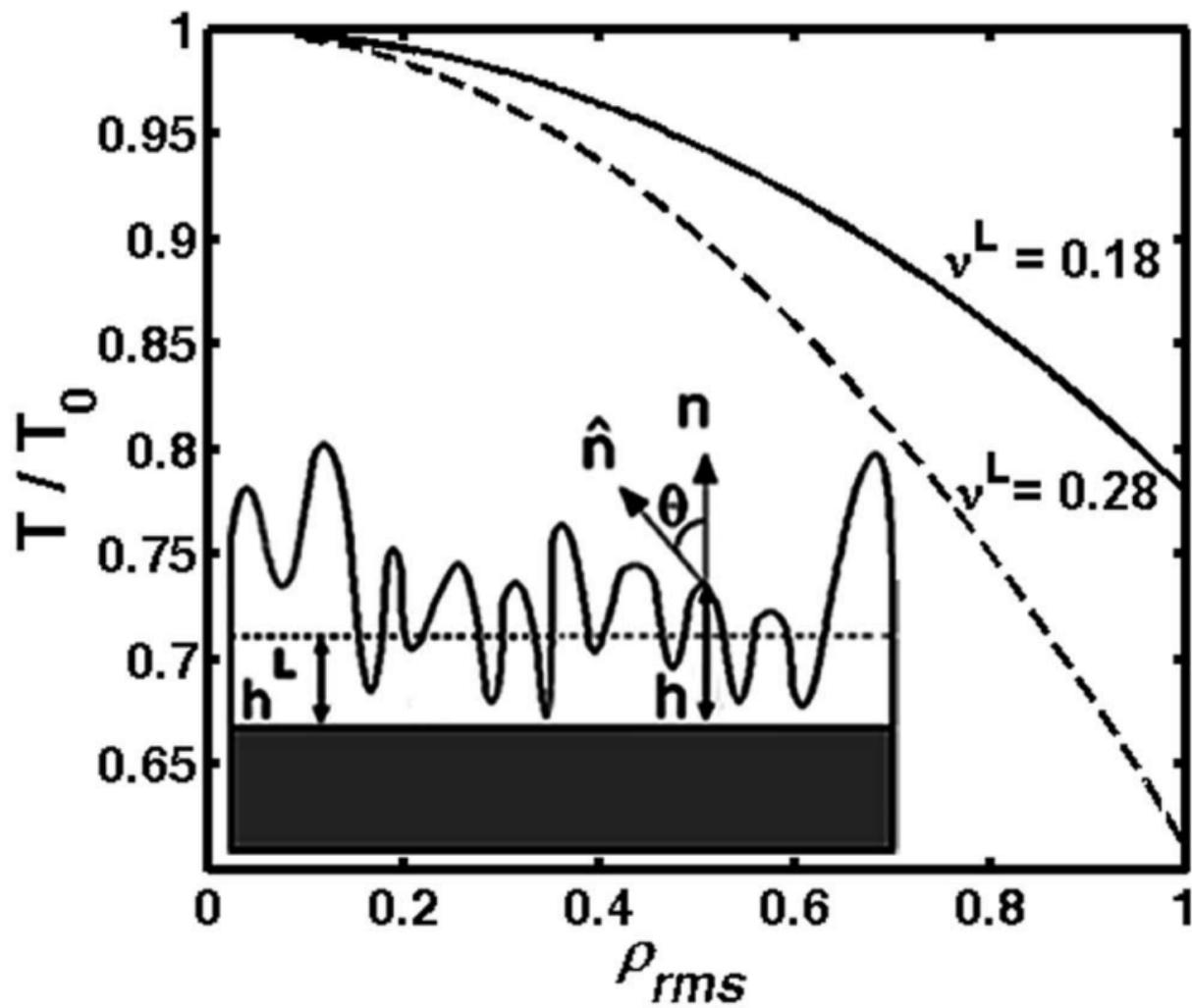

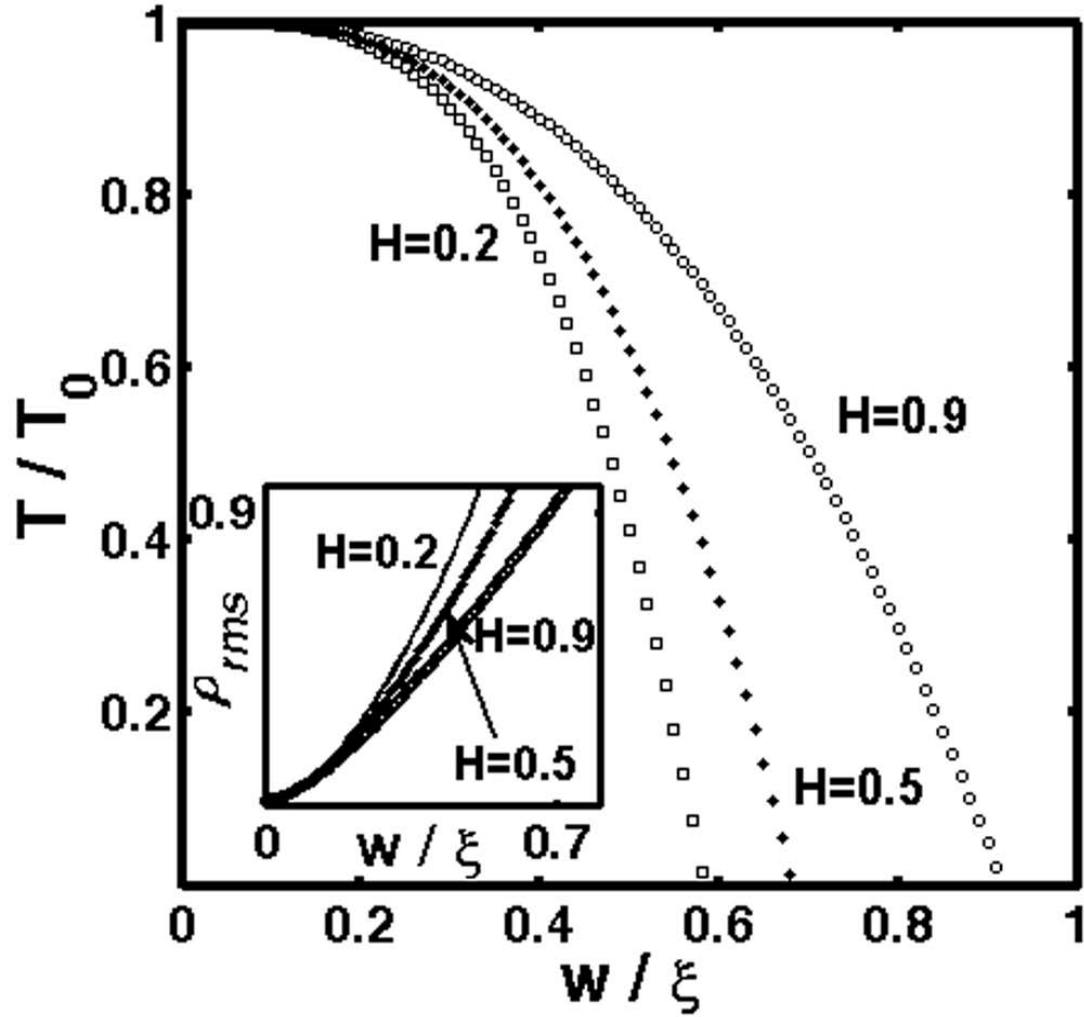

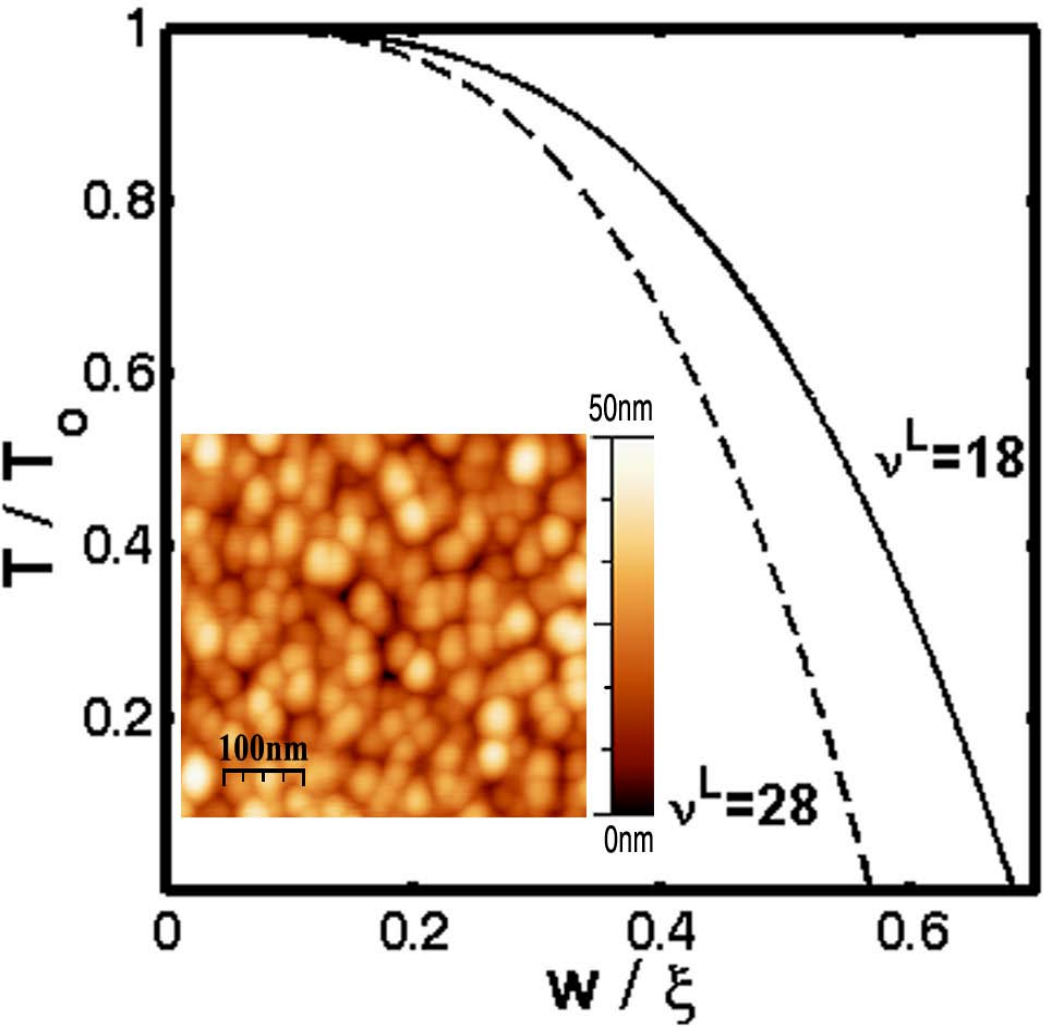